\documentclass[doublecol]{epl2}
\usepackage{amssymb}
\usepackage[usenames]{color}

\begin{document}

\title{Tightly bound gap solitons in a Fermi gas}
\shorttitle{Fermion gap solitons}

\author{Sadhan K. Adhikari\inst{1} \and Boris A. Malomed\inst{2}}
\institute{
\inst{1}{Instituto de F\'{\i}sica Te\'{o}rica, UNESP -- S\~{a}o Paulo
State
University, 01.405-900 S\~{a}o Paulo, S\~{a}o Paulo, Brazil }\\
\inst{2}{Department of Physical Electronics, School of Electrical
Engineering,\\
Faculty of Engineering, Tel Aviv University, Tel Aviv 69978,
Israel} }


\abstract{Within the framework of the mean-field-hydrodynamic
model of a degenerate Fermi gas (DFG), we study, by means of
numerical methods and variational approximation (VA), the
formation of fundamental gap solitons (FGSs) in a DFG (or in a BCS
superfluid generated by weak interaction between spin-up and
spin-down fermions), which is trapped in a periodic
optical-lattice (OL) potential. An effectively one-dimensional
(1D) configuration is considered, assuming strong transverse
confinement; in parallel, a proper 1D model of the DFG (which
amounts to the known quintic equation for the Tonks-Girardeau gas
in the OL) is considered too. The FGSs found in the first two
bandgaps of the OL-induced spectrum ({ unless they are}
very close to edges of the gaps) feature a ({\it tightly-bound})
shape, being essentially confined to a single cell of the OL. In
the second bandgap, we also find antisymmetric tightly-bound {\it
subfundamental solitons} (SFSs), with zero at the midpoint. The
SFSs are also confined to a single cell of the OL, but, unlike the
FGSs, they are unstable. The predicted solitons, consisting of
$\sim 10^4 - 10^5$ atoms, can be created by available experimental
techniques in the DFG of $^6$Li atoms.}

\maketitle

\textit{Introduction:} Matter-wave solitons were created as localized
nonlinear excitations in Bose-Einstein condensates (BECs) of attractively
interacting $^{7}$Li \cite{Li-soliton} and $^{85}$Rb \cite{Rb-soliton} atoms
loaded in a cigar-shaped trap. In both cases, the interaction was 
switched
from repulsion to attraction by applying magnetic field near the Feshbach
resonance \cite{Feshbach,bffesh}). In the absence of axial confinement, the
solitons can propagate freely in the axial direction. This was followed by
the creation of \textit{gap solitons}, GSs (formed by a few hundred atoms of
$^{87}$Rb) in a self-repulsive BEC loaded in a cigar-shaped trapped combined
with an optical lattice (OL) \cite{Markus}, which was created as the
interference pattern by counter-propagating laser beams (see also review 
\cite{Markus-review}). Theoretical description of the dilute BEC relies upon the
Gross-Pitaevskii equation (GPE), which provides for a remarkably accurate
description of various matter-wave patterns, including solitons 
\cite{BECsolitons}. In particular, GSs exist at values of the chemical potential
that fall in finite gaps of the band spectrum of the linear problem, induced
by the periodic OL potential. In that case, the system possesses a 
\textit{negative effective mass}, which allows the formation of solitons in balance
with the repulsive nonlinearity \cite{GSprediction}.

The quest for solitons in degenerate Fermi gases (DFGs), which are also
available to experiments \cite{Jin}, is a new challenge. Quasi-soliton
excitations in DFGs formed by noninteracting fermions were analyzed in refs.
\cite{Quasisoliton(Fermi-nonint)}. Solitons were predicted in fermion-boson
mixtures \cite{BFsoliton(BBrepBFattr),Sadhan-BFsoliton} which feature strong
fermion-boson attraction. Currently, $^{6}$Li-$^{7}$Li \cite{LiLi}, 
$^{6}$Li-$^{23}$Na \cite{LiNa}, and $^{40}$K-$^{87}$Rb \cite{KRb} mixtures of boson
and fermion atoms are available, as well as binary DFGs, such as mixtures of
two different spin states in $^{40}$K \cite{K} and $^{6}$Li \cite{Li} gases.
Accordingly, solitons in a binary fermion gas with attraction between the
two species have been predicted \cite{Sadhan-FFsoliton}.

In this Letter, we aim at  predicting solitons of the gap type in a DFG 
trapped in
the OL, using a mean-field hydrodynamic (MFHD) model of the DFG, which
actually stems from an approximation of the Thomas-Fermi type 
\cite{static,dynamic}. While the MFHD equations do not grasp effects of the
multi-particle coherence, they provide a reasonably accurate description of
various macroscopic patterns in DFGs, including the formation of solitons
\cite{skcol} and miscibility-immiscibility transition in binary gases 
\cite{we}. As explained below, essentially the same equations may also apply to a
different setting, \textit{viz}., a BCS\ (Bardeen-Cooper-Schrieffer)
superfluid formed in the gas of fermions due to a weak attraction between
atoms with opposite polarizations of the spin; the attraction may be induced
by means of the Feshbach-resonance technique too.

We consider three versions of the 1D MFHD equation for the fermion
wave function. The most fundamental one is derived from the
underlying 3D Fermi distribution, assuming a quasi-1D
(cigar-shaped) trap with the radius of the transverse confinement,
$a_{\perp }$, larger than the de Broglie wavelength, $\lambda
_{F}$, of atoms on the Fermi surface. In this case, the MFHD
equation for wave function $\psi $ contains the self-repulsive
nonlinear term $|\psi |^{4/3}\psi $. In the opposite case of an
extremely tight transverse confinement, with $a_{\perp }\ll
\lambda _{F}$, the underlying Fermi distribution is
one-dimensional, leading to an equation with the repulsive quintic
term, $|\psi |^{4}\psi $, the same as in the version of the GPE
derived in ref. \cite{Kolomeisky} for the Tonks-Girardeau gas.
Solitons in the 1D equation combining the quintic term and the OL
potential were recently considered in refs. \cite{Alfimov}. One
may also consider an anisotropic trap, with $a_{\perp }\gtrsim
\lambda _{F}$ in one transverse direction and $a_{\perp }\ll
\lambda _{F}$ in the other. Then, the underlying 2D Fermi
distribution gives rise to an equation with the self-repulsive
cubic term, $|\psi |^{2}\psi $, formally the same as in the
ordinary BEC.

GS solutions to the GPE with the repulsive nonlinearity are usually assumed
to be loosely bound ones, featuring weak localization and wavy tails, see,
e.g., refs. \cite{Alfimov}. We are interested in a possibility to predict
GSs of a different kind, \textit{tightly-bound} ones, localized practically
in a single cell of the OL (cf. refs. \cite{Thawatchai}, \cite{Hidetsugu},
and \cite{GMM}, where both loosely and tightly bound GSs were investigated
in BEC\ models). We demonstrate that this is the case indeed, except when
the chemical potential is very close to edges of the bandgap (in that case,
GSs {start to develop} wavy tails, signalizing a 
transition
to delocalized states in the Bloch bands). We report families of symmetric
(even) \textit{fundamental } \textit{gap solitons} (FGSs), originating in
the first bandgap and continuing into the second gap, which feature a
waveform with a single maximum, and antisymmetric (odd) \textit{
subfundamental } gap solitons (SFSs, the name borrowed from ref. 
\cite{Thawatchai}), that emerge in the second bandgap, with two maxima in the
density profile and a zero between them, all squeezed into a single OL cell.

Below, we study these solitons by means of numerical simulations and
variational approximation (VA) \cite{VA}. The VA, based on the Gaussian
ansatz, predicts the FGS families, in terms of the dependence between the
effective nonlinearity and chemical potential, in the first two bandgaps
with a surprisingly high accuracy, if compared to numerical results. SFSs
are predicted too, although less accurately, in the second bandgap, by means
of another (antisymmetric) ansatz. We also report examples of stable
symmetric and antisymmetric bound states of the FGSs.

\textit{The models} We derive the MFHD equation for the fermion
wave function, $\Psi $, as an extension of the static Thomas-Fermi
distribution of the atomic density, $n=|\Psi |^{2}$, which is 
\cite{static,dynamic}
\begin{equation}
V(\mathbf{r})+V_{\mathrm{eff}}^{\mathcal{D}}=\mu ,  \label{1}
\end{equation}
with $\mu $ the chemical potential, $V(\mathbf{r})$ the external trapping
potential, and $V_{\mathrm{eff}}^{\mathcal{D}}$ an effective nonlinear
potential in the $\mathcal{D}$-dimensional space accounting for the Fermi
pressure. For $\mathcal{D}\equiv $ 3D, 2D, and 1D, the role of 
$V_{\mathrm{eff}}^{(\mathcal{D})}$ is actually played by the local Fermi energy, $
\varepsilon _{\mathrm{F}}$, expressed in terms of local atomic density $n_{
\mathcal{D}}$, which is defined as per the dimension \cite{dynamic}. This
yields $\left( 2m/\hbar ^{2}\right) V_{\mathrm{eff}}^{(\mathcal{D})}=(6\pi
^{2}n_{3\mathrm{D}})^{2/3},4\pi n_{2\mathrm{D}},$ and $(\pi n_{1\mathrm{D}
})^{2},$ respectively, with $m$ the atom mass. The MFHD equation for
stationary problems is derived by treating relation (\ref{1}) as one which
defines the effective Thomas-Fermi Hamiltonian in terms of 
$V_{\mathrm{eff}}^{\mathcal{
D}}$. Augmenting the Hamiltonian with the kinetic-energy term and applying
it to the wave function, one arrives at the stationary equation 
\cite{dynamic},
\begin{equation}
-\frac{\hbar ^{2}}{2m}\nabla _{\mathcal{D}}^{2}\Psi +[V(\mathbf{r})+V_{
\mathrm{eff}}^{\mathcal{D}}]\Psi =\mu \Psi .  \label{2}
\end{equation}
A dynamic version of eq. (\ref{2}) is \cite{dynamic}
\begin{equation}
-\frac{\hbar ^{2}}{2m}\nabla _{\mathcal{D}}^{2}\psi +[V(\mathbf{r})+V_{
\mathrm{eff}}^{\mathcal{D}}]\psi =i\hbar \frac{\partial \psi }{\partial t},
\label{3}
\end{equation}
with$\psi (\mathbf{r},t)=e^{-i\mu t/\hbar }\Psi (\mathbf{r})$, although this
generalization is formal, as the above-mention derivation does not define
the phase of the fermion wave function. Nevertheless, the dynamical
equation, if it is treated as a phenomenologically postulated one, may yield
reasonable predictions for stability of static MFHD states in the DFG \cite
{dynamic,Sadhan-BFsoliton,Sadhan-FFsoliton}.

As mentioned above, dynamical equation (\ref{3}) can also be derived in a
different physical context, for a BCS superfluid formed by Cooper pairs of
fermions with opposite polarizations of the spin. Indeed, using the known
expression for the local energy density of the superfluid in 3D \cite{Yang},
$\mathcal{E}_{\mathrm{3D}}=(3/5)n_{3\mathrm{D}}\varepsilon _{\mathrm{F}}$
(recall $\epsilon _{\mathrm{F}}$ is the local Fermi energy), and deriving
the equation for the wave function of the Cooper pairs (alias the complex
order parameter, in terms of the Ginzburg-Landau approach) from the
Lagrangian density which includes term $\mathcal{E}_{\mathrm{3D}}$, one will
end up with a 3D equation that differs from the 3D version of eq. (\ref{3})
only by a numerical factor in front of $V_{\mathrm{eff}}^{\mathcal{D}}$. In
a similar fashion, using the local density of the BCS superfluid in 1D 
\cite{salasnich} and 2D settings, $\mathcal{E}_{\mathrm{1D}}=(1/3)n_{3\mathrm{D}
}\varepsilon _{\mathrm{F}}$ and $\mathcal{E}_{\mathrm{2D}}=(1/2)n_{2\mathrm{D
}}\varepsilon _{\mathrm{F}}$, one can derive the respective 1D and
2D equations, which differ from their counterparts (\ref{1D}) and
(\ref{2D}) written below only by numerical coefficients in front
of the nonlinear terms.

In the case of the ordinary cigar-shaped waveguide with strong harmonic
confinement in the transverse plane, eq. (\ref{3}) and its static version,
eq. (\ref{2}), can be reduced to the 1D form by means of the well-known
substitution \cite{Luca}, $\psi (x,y,z,t)=\phi (x,t)\exp \left[ -i\omega
_{\perp }t-\left( y^{2}+z^{2}\right) /\left( 2a_{\perp }^{2}\right) \right]
, $with the second multiplier representing the ground state of the
transverse harmonic oscillator, $\omega _{\perp }$ is the respective
frequency, and $a_{\perp }=\sqrt{\hbar /(m\omega _{\perp })}$. The
substitution of this ansatz in eq. (\ref{3}) for $\mathcal{D}=$ 3D and
averaging in the transverse plane yield the effective 1D equation,
\begin{equation}
i\hbar \phi _{t}=-\frac{\hbar ^{2}}{2m}\phi _{xx}+\frac{3\hbar ^{2}}{10m}
\left[ 6\pi ^{2}\left\vert \phi \right\vert ^{2}\right] ^{2/3}\phi -\epsilon
\cos \left( \frac{4\pi }{\lambda }x\right) \phi ,  \label{5}
\end{equation}
where the OL potential with strength $\epsilon $ and period $\lambda /2$ is
introduced. Equation (\ref{5}) is further rescaled by defining $\phi \equiv
\sqrt{{2N}/{\lambda }}a_{\perp }^{-1}\tilde{\phi},~t\equiv {\lambda ^{2}m}/{
\ (4\pi ^{2}\hbar )}\tilde{t},~x\equiv {\lambda }\tilde{x}/(2\pi )$, $
V_{0}=\epsilon m\lambda ^{2}/\left( {2\pi \hbar }\right) ^{2},$ where $N$ is
the number of atoms. The result is
\begin{equation}
i\phi _{t}=-(1/2)\phi _{xx}+g_{\mathrm{3D}}\left\vert \phi \right\vert
^{4/3}\phi -V_{0}\cos \left( 2x\right) \phi  \label{3D}
\end{equation}
(tildes are dropped here), with the wave function subject to normalization $
\mathcal{N}\equiv \int_{-\infty }^{+\infty }\left\vert \phi (x)\right\vert
^{2}dx=1,$ and effective strength of the nonlinearity and potential defined
as $g_{\mathrm{3D}}=({3}/{10})\left[ {3N\lambda ^{2}}/{(2\pi a_{\perp 
}^{2}})
\right] ^{2/3}$ (subscript $\mathrm{3D}$ refers to the derivation
of the equation from the 3D Fermi distribution), $V_{0}=m\epsilon
\left( \lambda /2\pi \hbar \right) ^{2}$. Similarly, the following
normalized 1D equations can be derived from the underlying 2D and
1D Fermi distributions:
\begin{eqnarray}
i\phi _{t} &=&-\frac{1}{2}\phi _{xx}+g_{\mathrm{2D}}\left\vert \phi
\right\vert ^{2}\phi -V_{0}\cos \left( 2x\right) \phi ,  \label{2D} \\
i\phi _{t} &=&-\frac{1}{2}\phi _{xx}+g_{\mathrm{1D}}\left\vert \phi
\right\vert ^{4}\phi -V_{0}\cos \left( 2x\right) \phi ,  \label{1D}
\end{eqnarray}
with $g_{\mathrm{2D}}=\lambda N/\left( \sqrt{\pi }a_{\perp }\right) ,~g_{
\mathrm{1D}}=\pi ^{2}N^{2}/2$.

Stationary solutions to eqs. (\ref{3D}), (\ref{2D}), and (\ref{1D}) are
looked for in the usual form, $\phi (x,t)=e^{-i\mu t}u(x)$, with real
function $u$ obeying the equation
\begin{equation}
\mu u+(1/2)u^{\prime \prime }-gu^{\aleph _{\mathcal{D}}}+V_{0}\cos \left(
2x\right) u=0,  \label{phi}
\end{equation}
where $\aleph _{\mathcal{D}}=7/3,3,5$, respectively, for $\mathcal{D}=$ 3D,
2D, 1D. In fact, the static 1D equations (\ref{phi}) for the real wave
function have the straightforward physical meaning, within the framework of
the MFHD description, while their dynamic counterparts for the complex wave
function are more formal ones, as mentioned above. Nevertheless, the
dynamical equations are quite useful, as their direct simulations converge
to real stationary states that represent numerically exact solutions to the
static equations (see below). Besides that, the dynamical equations make
sense (similar to the time-dependent Ginzburg-Landau equations) if applied,
as mentioned above, to the BCS superfluid. Our main objective is to find a
family of FGS solutions to eq. (\ref{phi}) with $\mu $ falling in the first
two finite bandgaps of the linear spectrum.

\textit{Variational approximation} Equation (\ref{phi}) and the above
normalization condition, $\mathcal{N}=1$, can be derived as the variational
equations \cite{VA}, $\delta L/\delta u=\partial L/\partial \mu =0$, from
the Lagrangian,
\begin{eqnarray}
L &=&\int_{-\infty }^{+\infty }\left[ \mu u^{2}-\frac{1}{2}\left( u^{\prime
}\right) ^{2}-\frac{2g}{\aleph _{\mathcal{D}}+1}u^{\aleph _{\mathcal{D}
}+1}\right.  \nonumber \\
&&\left. +V_{0}\cos (2x)\cdot u^{2}\right] dx-\mu .  \label{L}
\end{eqnarray}
To apply the VA, we use a simple Gaussian ansatz \cite{VA} ,
\begin{equation}
u(x)=\pi ^{-1/4}\sqrt{\mathcal{N}/W}\exp \left[ -x^{2}/\left( 2W^{2}\right) 
\right] ,  \label{Gauss}
\end{equation}
where variational parameters are the soliton's norm and width, $\mathcal{N}$
and $W$ (in addition to $\mu $). The substitution of this ansatz in
Lagrangian (\ref{L}) yields
\begin{eqnarray}
L &=&\mu \left( \mathcal{N}-1\right) -\frac{\mathcal{N}}{4W^{2}}+V_{0}
\mathcal{N}Ne^{-W^{2}}  \nonumber \\
&-&\frac{2^{3/2}g}{\pi ^{\aleph _{\mathcal{D}}/2}\left( \aleph _{\mathcal{D}
}+1\right) ^{3/2}}\frac{\mathcal{N}^{\left( \aleph _{\mathcal{D}}+1\right)
/2}}{W^{\left( \aleph _{\mathcal{D}}-1\right) /2}}.  \label{LGauss}
\end{eqnarray}
The corresponding variational equations, $\partial L/\partial \mu =\partial
L/\partial W=\partial L/\partial \mathcal{N}=0$, lead to $\mathcal{N}=1$ (as
expected), and
\begin{equation}
1+\frac{2^{3/2}\left( \aleph _{\mathcal{D}}-1\right) g}{\pi ^{\left( \aleph
_{\mathcal{D}}-1\right) /4}\left( \aleph _{\mathcal{D}}+1\right) ^{3/2}}
W^{\left( 5-\aleph _{\mathcal{D}}\right) /2}=4V_{0}W^{4}e^{-W^{2}},
\label{WGauss}
\end{equation}
\begin{equation}
\mu =\frac{1}{4W^{2}}+\frac{\sqrt{2}g}{(\pi W^{2})^{\left( \aleph _{\mathcal{
\ D}}-1\right) /4}\sqrt{\aleph _{\mathcal{D}}+1}}-V_{0}e^{-W^{2}}.
\label{muGauss}
\end{equation}
Then, eqs. (\ref{WGauss}) and (\ref{muGauss}) were solved numerically.

\textit{Results} Simulations of eqs. (\ref{3D}), (\ref{2D}), and (\ref{1D})
were carried out by dint of a real-time integration method based on the
Crank-Nicholson discretization scheme, as elaborated in ref. \cite{sk1}. The
equations were discretized using time and space steps $0.0005$ and $0.025$,
respectively, in domain $-20<x<20$. To find FGSs, the simulations started
with an initial configuration chosen as the ground state, $\phi (x)=(\sqrt{2}
c/\pi )^{1/4}{\exp [-x^{2}\sqrt{c/2}]}$, of the linear oscillator with
potential $cx^{2}$ (with $c\gg 1$, typically 5 to 100). The OL potential was
slowly introduced in the course of the simulations. The strong harmonic
potential squeezes the localized state into a single cell of the OL. After
obtaining a stationary bound state in the nonlinear equation containing the
combined OL and harmonic potential, the latter one was slowly switched off.
The simulations were run until a well-defined stationary shape of the
soliton was established. A similar approach to the creation of GSs in BEC
was proposed in ref. \cite{Canberra}. To generate SFS solutions (see below),
the initial state was taken as the first excited state of the
above-mentioned harmonic potential, instead of its ground state. We present
results for a characteristic value of the OL strength, $V_{0}=5$.

\begin{figure}[tbp]
\begin{center}
{\includegraphics[width=\linewidth]{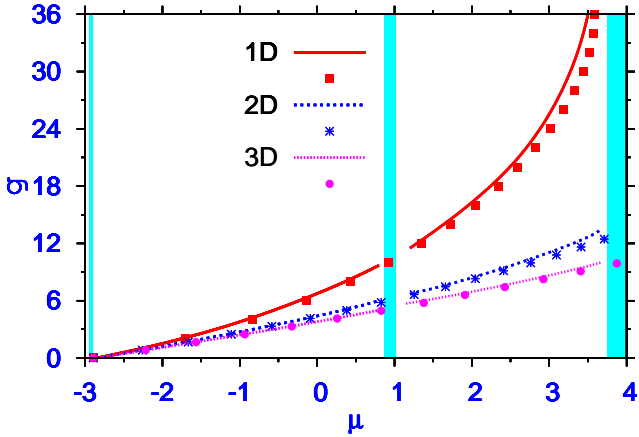}}
\end{center}
\caption{Numerical (continuous curve) and variational (chain of symbols)
results for nonlinearity $g$ versus chemical potential $\protect\mu $, for
fundamental gap solitons in the first and second finite bandgaps of periodic
potential $-V_{0}\cos (2x)$, (all results reported in this Letter with 
$V_{0}=5$,) in the models based on eqs. 
(\protect\ref{1D}), (\protect\ref{2D}), and (\protect\ref{3D})
(``1D", ``2D", and ``3D", respectively, which refers to the
dimension of the underlying Fermi distribution, from which the
equation is derived). The shaded areas represent the Bloch bands.}
\label{Fig1}
\end{figure}

Families of FGS solutions are characterized by the corresponding
dependences, $g_{\mathcal{D}}(\mu )$. In fig. \ref{Fig1}, they are displayed
as found from the numerical solution of the 1D, 2D, and 3D models, along
with the same dependences as predicted, for the three models, by the VA.

In fig. 2 we display typical profiles of the solitons. It is obvious that
the FGSs, unless taken too close to bandgap edges, are compact objects,
primarily trapped in a single cell of the OL. There also exist more loosely
organized localized states, which occupy several cells \cite{Alfimov}, that
we do not consider here. Figure 2 demonstrates that the VA produces a very
good fit not only to the $g(\mu )$ plots for the entire soliton families
(see fig. 1), but also to FGS profiles in the two lowest finite gaps, except
very close to edges of the gaps, where the solitons develop undulate tails,
that simple ansatz (\ref{Gauss}) is unable to reproduce [see panels in fig.
2 pertaining to $g=0.01$].
{In principle, GSs with conspicuous tails may be
approximated by a more complex ansatz, which combines the Gaussian
and function $\cos x$; however, variational equations generated by
such an ansatz are very cumbersome, and, in the end, the extended
ansatz does not approximate the entire tail, but only its
secondary peaks which are closest to the central one \cite{GMM}.
For this reason, we do not try to apply that ansatz here.}

\begin{figure}[tbp]
{\includegraphics[width=\linewidth]{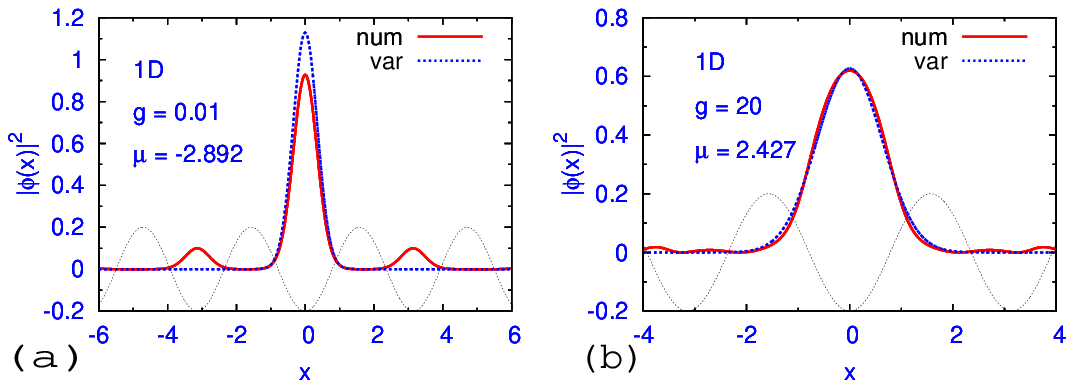}} {\includegraphics[width=
\linewidth]{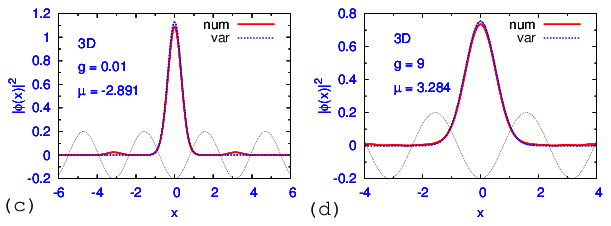}}
\caption{Typical shapes of the fundamental gap solitons in the model derived
from the 1D Fermi distribution, Eq. (\ref{1D}), for (a) 
$g_{1\mathrm{D}}=0.01$, 
($\protect\mu =-2.892$),
(b) $g_{1\mathrm{D}}
=
20$ ($\protect\mu =2.427$), and from the 3D distribution, 
Eq. (\ref{3D}), for (c) $
g_{3\mathrm{D}}=0.01$, ($\protect\mu =-2.891$), and (d) 
$g_{3\mathrm{D}}=9$, ($\protect\mu
=3.284$ ). Shapes produced by the numerical solution and
variational approximation are labeled as ``num" and ``var". The
panels pertaining to $g_{3\mathrm{D}}=g_{1\mathrm{D}}
=0.01$ display the 
solitons very close to
the left edge of the first bandgap, while the panels for  $g_{1\mathrm{D}}
=
20$ 
and $g_{3\mathrm{D}}=9$ give typical examples found inside the second 
bandgap
(inside the first gap, the shape of the fundamental solitons is
quite similar). The thin sinusoidal line (in this and following
figures) depicts the profile of the underlying OL potential. }
\label{Fig2}
\end{figure}

{It is relevant to mention that as the chemical potential 
moves closer to  the Bloch band the wave form of the GSs develop a
structure in space similar to Wannier functions, which are
spatially localized linear combinations of Bloch functions. Figure
\ref{Fig2} 
(a) shows the wave form near the edges of the gap. The undulate 
tails in this figure signal the proximity to a Bloch band, where the
wave form  
will turn into a periodic structure. However, as said above, such wave 
forms cannot be fitted to 
Gaussian shapes and we do not intend to study them in
detail here.}

\begin{figure}[tbp]
\begin{center}
\includegraphics[width=\linewidth,clip]{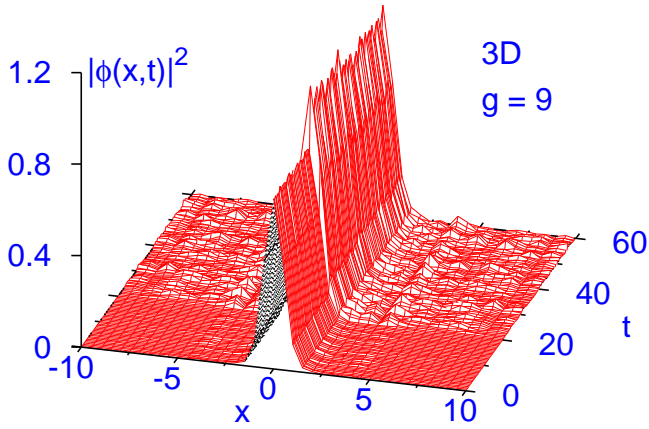}
\end{center}
\caption{An example of stable evolution of a fundamental gap soliton
solution in the 3D model [eq. (\protect\ref{3D})], for $g_{3\mathrm{D}}=9$ 
of Fig. \ref{Fig2} (d).
To introduce a perturbation, at $t=20$, the wave form was suddenly
multiplying by $1.2$: $\phi(x,t) \to 1.2\times \phi(x,t)$. The so 
perturbed soliton (as well as ones subjected to
different arbitrary perturbations) remained stable as long as the
simulations were run.}
\label{fig3N}
\end{figure}

Within the framework of eqs. (\ref{3D}) - (\ref{1D}), the
stability of the FGSs was tested by subjecting them to relatively
strong initial perturbations (as said above, the dynamical
equations have direct meaning in the application to the BCS
superfluid). It has been found that the entire families of these
solutions are stable, in all the three above-mentioned models, see
a typical example in Fig. \ref{fig3N}. In the case displayed in
this figure, after the stationary FGS of the model based on the 3D
Fermi distribution was obtained (for $g_{3\mathrm{D}}=9$), we
replaced, at $t=20$, the stationary wave form $\phi (x)$ by a
perturbed one, $1.2\phi (x)$, and continued the simulation. The
resultant solution demonstrates persistent oscillations, and no
sign of degradation.

The usual GPE, with the repulsive cubic nonlinearity and OL potential, i.e.,
eq. (\ref{2D}), gives rise to antisymmetric SFSs, the family of which starts
at the left edge of the second finite bandgap \cite{Thawatchai}. A
characteristic feature of the SFSs is that two maxima of the density, $
\left\vert \phi (x)\right\vert ^{2}$, are located inside a single site of
the periodic potential (i.e., this antisymmetric soliton as a whole is
confined to a single site). To look for SFSs in the present models, we
applied the VA based on the modified Gaussian ansatz (cf. eq. (\ref{Gauss}
)),
\begin{equation}
u(x)=\pi ^{-1/4}\left( \sqrt{2\mathcal{N}}/W^{3/2}\right) x\exp \left(
-x^{2}/\left( 2W^{2}\right) \right) ,  \label{Gauss1}
\end{equation}
where $N$ and $W$ have the same meaning as in eq. (\ref{Gauss}). The
substitution of ansatz (\ref{Gauss1}) in Lagrangian (\ref{L}) yields
\begin{eqnarray}
L &=&\mu \left( \mathcal{N}-1\right) -\frac{3\mathcal{N}}{4W^{2}}+{V}_{0}
\mathcal{N}e^{-W^{2}}(1-2W^{2})  \nonumber \\
&-&\frac{2^{\aleph _{\mathcal{D}}+5/2}g\Gamma \left( 1+\aleph _{\mathcal{D}
}/2\right) }{\pi ^{(\aleph _{\mathcal{D}}+1)/4}\left( \aleph _{\mathcal{D}
}+1\right) ^{\left( \aleph _{\mathcal{D}}+4\right) /2}}\frac{\mathcal{N}
^{\left( \aleph _{\mathcal{D}}+1\right) /2}}{W^{\left( \aleph _{\mathcal{D}
}-1\right) /2}},  \label{LGauss1}
\end{eqnarray}
and the variational equations following from here are $\mathcal{N}=1$ and
\begin{eqnarray}
1 &+&\frac{\aleph _{\mathcal{D}}-1}{3}\frac{2^{(2\aleph _{\mathcal{D}
}+5)/2}gW^{(5-\aleph _{\mathcal{D}})/2}}{\pi ^{(\aleph _{\mathcal{D}
}+1)/4}\left( \aleph _{\mathcal{D}}+1\right) ^{(4+\aleph _{\mathcal{D}})/2}}
\Gamma (\aleph _{\mathcal{D}}/2+1)  \nonumber \\
&=&(4/3)V_{0}W^{4}e^{-W^{2}}(3-2W^{2}),  \label{WGauss1}
\end{eqnarray}
\begin{eqnarray}
\mu  &=&\frac{3}{4W^{2}}+\frac{2^{(2\aleph _{\mathcal{D}}+3)/2}g}{\pi
^{(\aleph _{\mathcal{D}}+1)/4}\left( \aleph _{\mathcal{D}}+1\right)
^{(2+\aleph _{\mathcal{D}})/2}}\frac{\Gamma (\aleph _{\mathcal{D}}/2+1)}{
W^{(\aleph _{\mathcal{D}}-1)/2}}  \nonumber \\
&-&{V_{0}}e^{-W^{2}}(1-2W^{2}).  \label{muGauss1}
\end{eqnarray}

\begin{figure}[tbp]
\begin{center}
{\includegraphics[width=.8\linewidth]{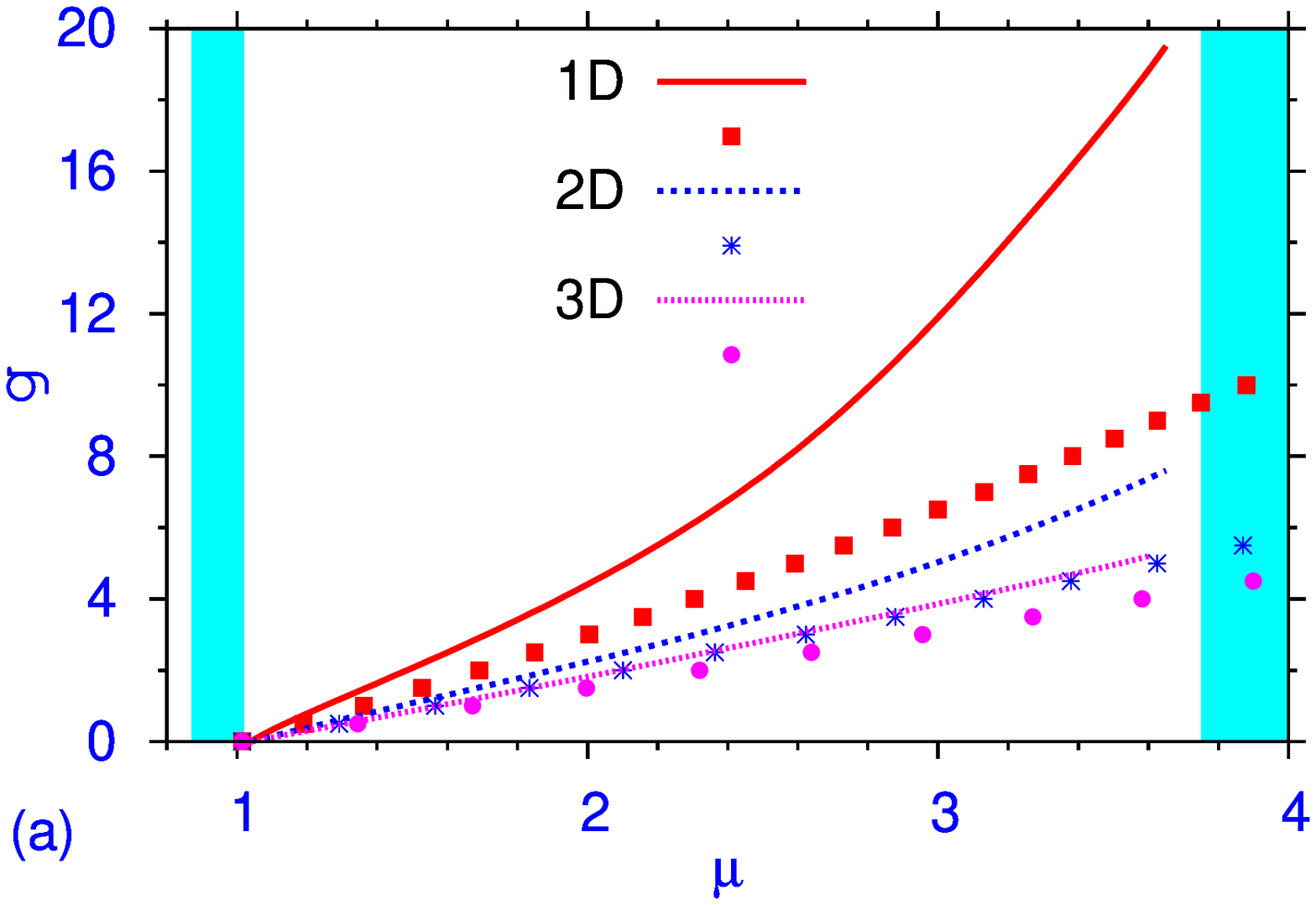}} {
\includegraphics[width=.8
\linewidth]{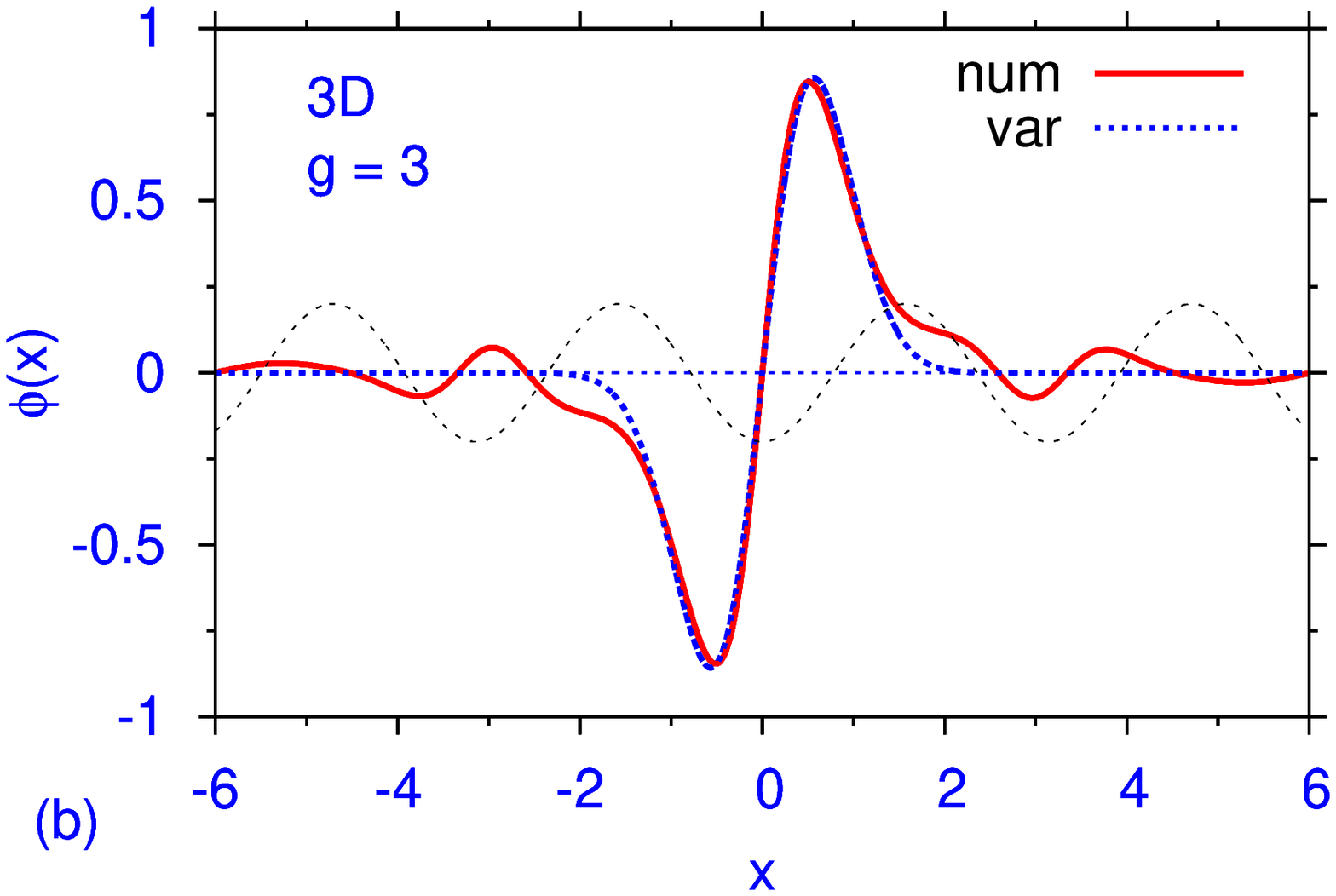}}
\end{center}
\caption{(a) The same as in fig. \protect\ref{Fig1}, but for families of
subfundamental gap solitons. (b) A typical example of the subfundamental
soliton, obtained from the numerical solution of eq. (\protect\ref{3D}) (for
$g_{3\mathrm{D}}=3$), and its counterpart predicted by the VA. }
\label{Fig4}
\end{figure}

Numerical solution of eqs. (\ref{WGauss1}) and (\ref{muGauss1}) gives rise
to SFS families in the second bandgap. The comparison of the corresponding
characteristics $g(\mu )$ with their numerically found counterparts is
presented in fig. \ref{Fig4}(a). The accuracy of the VA predictions for the
SFS families is worse than for the FGS solutions; nevertheless, the
prediction is qualitatively correct.

Although the SFS families were found from direct simulations of eqs. 
(\ref{3D}), (\ref{2D}), and (\ref{1D}), hence they are, at least, quasi-stable
solutions, longer simulations reveal their instability (not shown here).
Similar to what was found in ref. \cite{Thawatchai} in the GPE with the
cubic nonlinearity, which coincides with eq. (\ref{2D}), in all the three
models considered here, the instability tends to transform the SFS
(double-humped) solitons into a stable FGS (single-humped).

\begin{figure}[tbp]
\begin{center}
{\includegraphics[width=.8\linewidth]{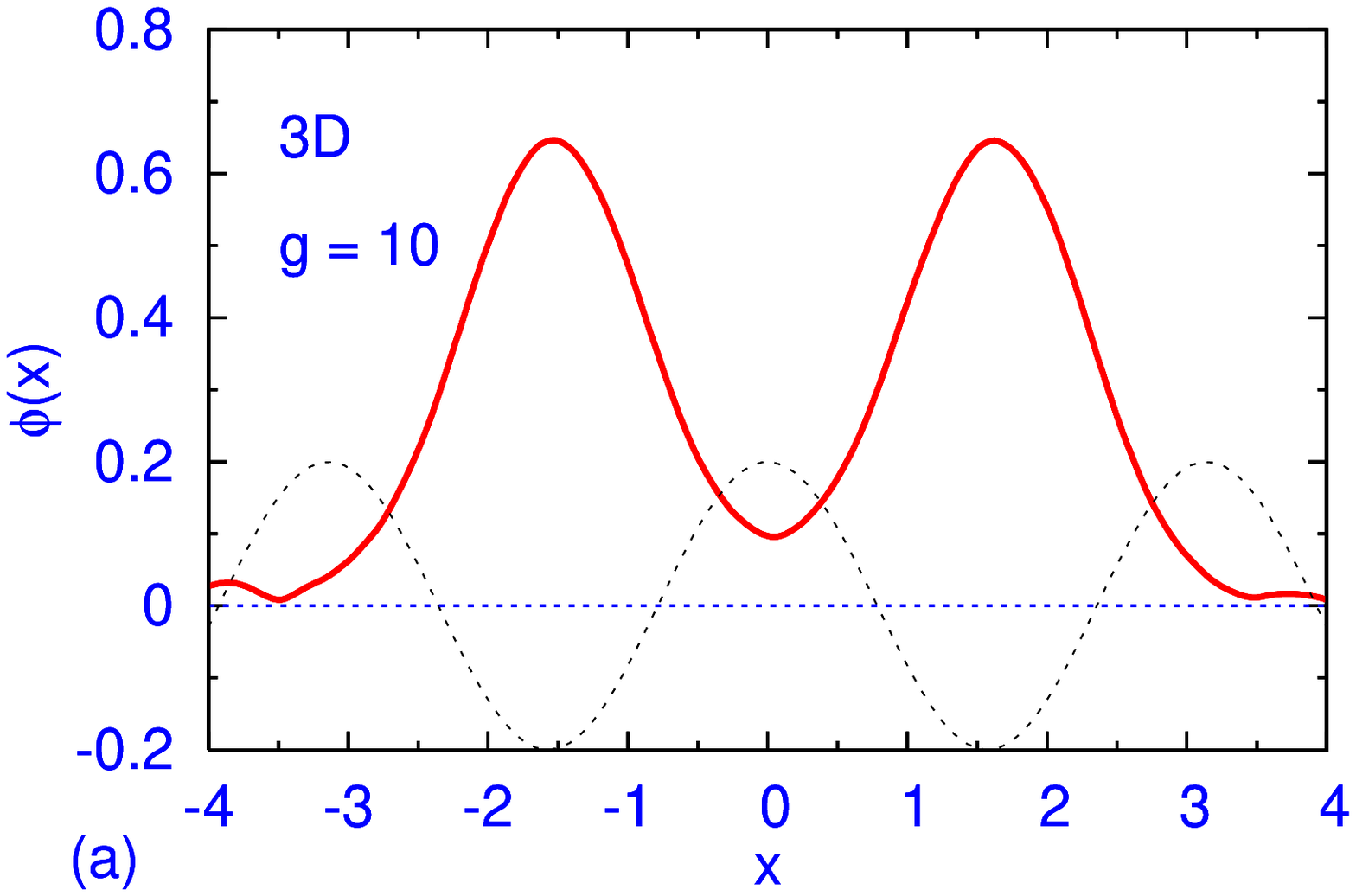}} {
\includegraphics[width=.8
\linewidth]{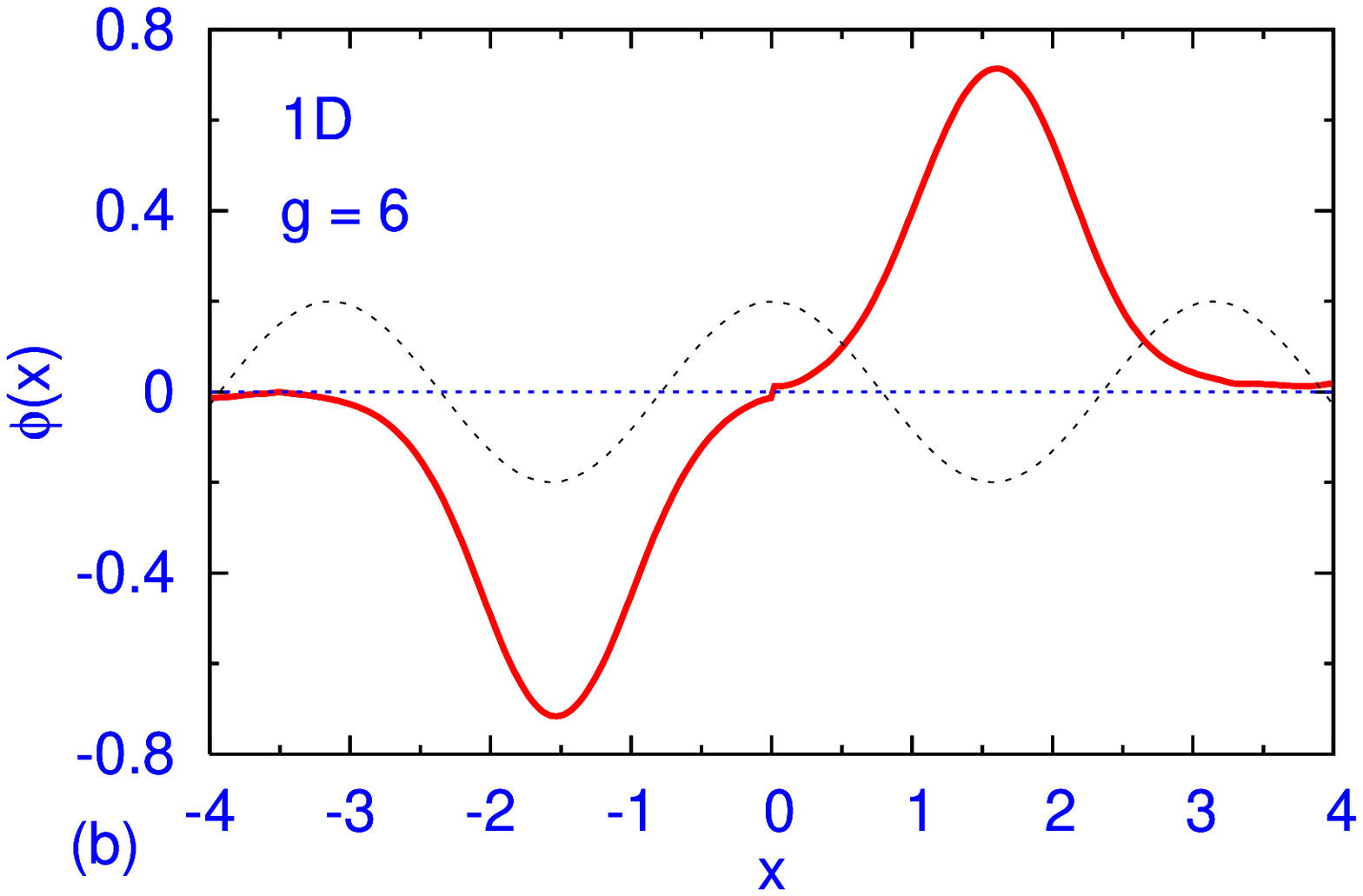}}
\end{center}
\caption{Examples of stable symmetric (a) and antisymmetric (b) bound states
of fundamental gap solitons in the models based on the 3D [Eq. (\ref{3D})]
and the 1D  [Eq. (\ref{1D})]
Fermi
distributions, respectively, for (a) $g_{3\mathrm{D}}=10$ and (b) $g_{1
\mathrm{D}}=6$.}
\label{Fig6}
\end{figure}

In addition to the SFSs, symmetric and antisymmetric (``twisted",
in the latter case) bound states of FGSs were found too. Typical
examples of such stable bound states in the 3D and 1D models are
shown in figs. \ref{Fig6}(a) and (b), respectively. The
antisymmetric state is formally similar to the SFS, but a
difference is that the density maxima of the bound state are
located in different OL sites. Direct simulations (not shown here)
clearly demonstrate that both symmetric and antisymmetric bound
states are stable one (on the contrary to the SFS solutions).

\textit{Conclusion} In the three MFHD models, based on eqs. (\ref{1D}), 
(\ref{2D}), and (\ref{3D}), we have considered, by means of numerical simulations
and VA (variational approximation), the generation of stable fundamental and
unstable subfundamental gap solitons (FGSs and SFSs, respectively) in the
quasi-1D degenerate Fermi gas loaded in a periodic OL potential. The same
models apply to the BCS superfluid trapped in the OL. In most cases, both
the FGSs and SFSs are confined, essentially, to a single cell of the
lattice. The VA provides a very good fit to the FGSs, and a qualitatively
correct approximation for the SFSs. Stable symmetric and antisymmetric bound
states of the FGSs were also found. It was demonstrated that transport of
the FGSs, without much distortion, is possible by a slowly moving OL.

Experimental realization of the FGSs in degenerate Fermi gases seems quite
feasible. To this end, the gas should be loaded in a cigar-shaped trap
combined with the periodic axial potential, as was done to create gap
solitons in the BEC \cite{Markus,Markus-review}. The experiment may be
started with an additional strong parabolic potential acting in the axial
direction, to prepare a strongly localized state, that may have a good
chance to self-trap into a tightly bound soliton while the extra potential
is gradually switched off. An estimate for the $^{6}$Li gas, with typical
values of physical parameters, predicts that the FGSs based on the 3D Fermi
distribution, i.e., obeying eq. (\ref{3D}), may be created with $
10^{4}-10^{5}$ atoms per soliton.

\acknowledgments

The work of B.A.M. was supported, in a part, by Israel Science Foundation
through the grant No. 8006/03. The work of S.K.A. was supported, in a part,
by FAPESP and CNPq of Brazil.


\begin{thebibliography}{99}
\bibitem{Li-soliton} P\'{e}rez-Garc\'{\i}a V. M. \textit{et al., Phys. Rev. A
}, \textbf{57} (1998) 3837; Strecker K. E. \textit{et al., Nature,} \textbf{
417} (2002) 150; Khaykovich L. \textit{et al., Science}, \textbf{256} (2002)
1290; Cornish S. L. \textit{et al., Phys. Rev. Lett.,} \textbf{96} (2006)
170401.

\bibitem{Rb-soliton} Cornish S. L., Thompson S. T., and Wieman C. E.,
\textit{Phys. Rev. Lett.} \textbf{96} (2006) 170401.

\bibitem{Feshbach} Inouye S. \textit{et al., Nature,} \textbf{392} (1998)
151.

\bibitem{bffesh} Zaccanti M. \textit{et al., Phys. Rev. A,} \textbf{74}
(2006) 041605(R); Ospelkaus S. \textit{et al., Phys. Rev. Lett.}, \textbf{97
} (2006) 120403.

\bibitem{Markus} Eiermann B. \textit{et al., Phys. Rev. Lett.,} \textbf{92}
(2004) 230401.

\bibitem{Markus-review} Morsch O. and Oberthaler M., \textit{Rev. Mod. Phys.,
} \textbf{\ \ 78} (2006) 179.


\bibitem{BECsolitons} Strecker K. E. \textit{et al., New J. Phys.,} \textbf{5
} (2003) 73; Brazhnyi V. A. and Konotop V. V., \textit{Mod. Phys. Lett. B,}
\textbf{18} (2004) 627; Abdullaev F. Kh. {\ \textit{et al., Int. J. Mod.
Phys. B,}} \textbf{19} (2005) 3415.

\bibitem{GSprediction} Alfimov G. L. \textit{et al., Europhys. Lett.},
\textbf{58} (2002) 7; Baizakov B. B. \textit{et al., J. Phys. B} \textbf{35}
(2002) 51015; Louis P. J. Y. \textit{et al., Phys. Rev. A} \textbf{67}
(2003) 013602. 

\bibitem{Jin} DeMarco B. and Jin D. S., \textit{Science}, \textbf{285}
(1999) 1703; Minguzzi A. \textit{et al., Phys. Rep.,} \textbf{395} (2004)
223; Chen Q. J. \textit{et al.}, \textit{Phys. Rep.}, \textbf{412} (2005) 
1.

\bibitem{Quasisoliton(Fermi-nonint)} Karpiuk T. \textit{et al., Phys. Rev. A,
} \textbf{66} (2002) 023612; Witkowska E. and Brewczyk M., \textit{Phys.
Rev. A,} \textbf{72} (2005) 023606.

\bibitem{BFsoliton(BBrepBFattr)} Karpiuk T. \textit{et al., Phys. Rev. Lett.}
, \textbf{93} (2004) 100401; T. Karpiuk T. \textit{et al., Phys. Rev. A},
\textbf{73} (2006) 053602.

\bibitem{Sadhan-BFsoliton} Adhikari S. K., \textit{Phys. Rev. A,} \textbf{72}
(2005) 053608; Salasnich L. \textit{et al., Phys. Rev. A,}
{\bf 75} (2007) 023616; Adhikari S. K.,
{\it J. Phys. A,} {\bf 40} (2007) 2673.  

\bibitem{LiLi} Schreck F. \textit{et al., Phys. Rev. Lett.}, \textbf{87}
(2001) 080403 (2001); Truscott A. G. \textit{et al., Science,} \textbf{291}
(2001) 2570.

\bibitem{LiNa} Hadzibabic Z. \textit{et al., Phys. Rev. Lett.,} \textbf{88}
(2002) 160401.

\bibitem{KRb} Modugno G. \textit{et al., Science,} \textbf{297} (2002) 2240;
Roati G. \textit{et al., Phys. Rev. Lett.}, \textbf{89} (2002) 150403.

\bibitem{K} DeMarco B. and Jin D. S., \textit{Science,} \textbf{285} (1999)
1703.

\bibitem{Li} O'Hara K. M. \textit{et al., Science,} \textbf{298} (2002)
2179; Strecker K. E. \textit{et al., Phys. Rev. Lett.,} \textbf{91} (2003)
080406.

\bibitem{Sadhan-FFsoliton} Adhikari S. K., \textit{Eur. Phys. J. D,} \textbf{
40} (2006) 157; Adhikari S. K., \textit{J. Phys. A, } \textbf{40} (2007)
2673.



\bibitem{static} Amoruso A. \textit{et al., Eur. Phys. J. D,} \textbf{8}
(2000) 361; Roth R. and Feldmeier H., \textit{J. Phys. B,} \textbf{34}
(2001) 4629; Molmer K., \textit{Phys. Rev. Lett.,} \textbf{80} (2004) 1804;
Modugno M. \textit{et al., Phys. Rev. A,} \textbf{68} (2003) 043626; 
Jezek D. M. \textit{et al.,} \textit{Phys. Rev. A}, \textbf{70} (2004) 
043630.


\bibitem{dynamic} Capuzzi P. \textit{et al., Phys. Rev. A,} \textbf{67}
(2003) 053605; \textbf{68} (2003) 033605.


\bibitem{skcol} Adhikari S. K., \textit{New J. Phys.,} \textbf{8} (2006)
258; 
Adhikari S. K., \textit{Phys. Rev. A,} \textbf{70}(2004) 043617.


\bibitem{we} Adhikari S. K. and Malomed B. A., \textit{Phys. Rev. A},
\textbf{74} (2006) 053620;
Adhikari S. K. and Salasnich L.,
 \textit{Phys. Rev. A,} \textbf{75} (2007) 053603; 
Adhikari S. K., \textit{Phys. Rev. A,} \textbf{73} (2006) 043619.



\bibitem{Kolomeisky} Kolomeisky E. B. \textit{et al.}
, \textit{Phys. Rev. Lett}., \textbf{85} (2000) 1147.


\bibitem{Alfimov} Abdullaev F. Kh. and Salerno M., \textit{Phys. Rev. A}
\textbf{72} (2005) 033617; Alfimov G. L. \textit{et al., Phys. Rev. A,}
\textbf{56} (2007) 023624; 
Salerno M., {\it Phys. Rev. A}, {\bf{72}} (2005) 063602.


\bibitem{Thawatchai} Mayteevarunyoo T. and Malomed B. A., \textit{Phys. Rev.
A,} \textbf{74} (2006) 033616.

\bibitem{Hidetsugu} Sakaguchi H. and Malomed B. A., \textit{J. Phys. B},
\textbf{37} (2004) 1443; 2225;
Gubeskys A.,  Malomed B. A,, and  Merhasin I. M., {\it Phys.
Rev. A},
\textbf{73} (2006) 023607.


\bibitem{GMM} Gubeskys A., Malomed B. A., and Merhasin I. M., 
\textit{Stud. Appl.
Math.} \textbf{115} (2005) 255.

\bibitem{VA} P\'{e}rez-Garc\'{\i}a V. M. \textit{et al., Phys. Rev. A},
\textbf{56} (1997) 1424; 
Malomed B. A., in \textit{Progress in Optics}, vol. 43, p. 71 (ed. by E.
Wolf: North-Holland, Amsterdam, 2002).


\bibitem{Yang} Huang K. and Yang C. N.  \textit{Phys. Rev.}
\textbf{105} (1957) 767; Lee T. D.
and Yang C. N.  \textit{Phys. Rev.} \textbf{105} (1957) 1119.

\bibitem{salasnich} Manini N. and Salasnich L.  \textit{Phys. Rev. A} 
\textbf{71} 
(2005)
033625.


\bibitem{Luca} Salasnich L., Parola A., and Reatto L., \textit{Phys. Rev. A}%
, \textbf{65} (2002) 043614; \textbf{66} (2002), 043603.


\bibitem{sk1} Adhikari S. K. and Muruganandam P., \textit{J. Phys. B},
\textbf{35} (2002) 2831. ; Muruganandam P. and Adhikari S. K.,
\textit{J. Phys. B},
\textbf{36} (2003) 2501.

\bibitem{Canberra} Matuszewski M. \textit{et al}.,
, Phys. Rev. A \textbf{73} (2006) 063621.




\end{thebibliography}
\end{document}